\documentclass{sf2a-conf2019}
\usepackage{graphicx}
\usepackage{amsmath}
\usepackage{hyperref}
\usepackage[]{natbib}  
\usepackage{epstopdf}

\def\BibTeX{{\rm B\kern-.05em{\sc i\kern-.025em b}\kern-.08em
    T\kern-.1667em\lower.7ex\hbox{E}\kern-.125emX}}
\bibpunct{(}{)}{;}{a}{}{,}  

\begin{document}

\TitreGlobal{SF2A 2019}


\title{Automatic classification of K2 pulsating stars using Machine Learning techniques}

\runningtitle{Automatic classification of K2 pulsating stars using Machine Learning techniques}

\address{IRFU, CEA, Université Paris-Saclay, 91191 Gif-sur-Yvette, France }

\address{AIM, CEA, CNRS, Université Paris-Saclay, Université Paris Diderot, Sorbonne Paris Cité, 91191 Gif-sur-Yvette, France}

\address{Universidad de la Laguna, Dpto. De Astrof\'\i sica, 38205 La Laguna, Tenerife, Spain\\ $^4$ Instituto de Astrofísica de Canarias, 38200 La Laguna, Tenerife, Spain}

\author{A. Le Saux$^{1,2}$}

\author{L. Bugnet$^{1,2}$}

\author{S. Mathur$^{3,4}$}

\author{S. N. Breton$^{1,2}$}

\author{R. A. Garc\'\i a$^{1,2}$}




\setcounter{page}{237}


\maketitle


\begin{abstract}
The second mission of NASA's \textit{Kepler} satellite, K2, has collected hundreds of thousands of lightcurves for stars close to the ecliptic plane. This new sample could increase the number of known pulsating stars and then improve our understanding of those stars. For the moment only a few stars have been properly classified and published. In this work, we present a method to automaticly classify K2 pulsating stars using a Machine Learning technique called Random Forest. The objective is to sort out the stars in four classes: red giant (RG), main-sequence Solar-like stars (SL), classical pulsators (PULS) and Other. To do this we use the effective temperatures and the luminosities of the stars as well as the FliPer features, that measures the amount of power contained in the power spectral density. The classifier now retrieves the right classification for more than 80\% of the stars.
\end{abstract}



\section{Introduction}
Following the end of the original NASA’'s \textit{Kepler} mission \citep{Borucki10}, K2 \citep{Howell14} targeted more than 580,000 stars close to the ecliptic between campaign 1 and 20. Today, only a small fraction of these stars are properly classified attending to their pulsation properties. This is why in this project, the objective is to sort this huge amount of pulsating stars with an automatic Random Forest (RF) classifier. The RF classifier is a supervised ML algorithm that consists in an ensemble of decision trees. This method is suitable for star classification because it can deal quickly with a very large amount of data.
The difficulty when analyzing K2 data is the special running mode implying a correction of the orbit every six hours that needs to be properly corrected. As a result, the Signal-to-Noise Ratio (SNR) is degraded compared to the \textit{Kepler} main mission. Another issue that had to be faced is related to the Machine Learning (ML) method. Only a few pulsating-stars catalogs are published for the moment, hence it is difficult to form a training set to train the RF classifier in all the parameter space.
  
\section{Methods}

\subsection{ Data }
 An important part of the work when using ML techniques is to pre-process the data that will be used, particularly with classification algorithms. The objective is to differentiate between types of stars that could be very similar so values of the used features have to be as accurate as possible. In this work, we aim at classifying the stars by using different parameters such as the effective temperature, the luminosity and the amount of power contained in their Power Spectral Density (PSD) because this combination of parameters provide one of the best tools to obtain information about the nature of the pulsating stars. The PSDs are obtained by computing the Fourier transform of the lightcurves. 

Lightcurves used in the project were produced with the EVEREST pipeline developped by \citet{Luger16, Luger18}, which aimed at studying exoplanet transit. For stars exhibiting transits, the PSD is dominated by the harmonics of the transiting periods,which could bias the total amount of power. Therefore the stars will be found in the wrong category. Other EVEREST lightcurves show jumps, discontinuities and low-frequency trends that introduce power in the PSD (for more information on those effects see \citet{Garcia11, Garcia14}). To remove those side effects, the available lightcurves were processed with a filtering method inspired by the KASOC filter from \citet{Handberg14}. This filter was developped to optimize the data for asterosismic analysis by removing instrumental effects and planetary signals. First of all, it was noted that many light curves showed a general increasing or decreasing trend, which affects the calculated values of the PSD. To overcome this, a moving median filter was applied to these timeseries.  A light curve is stored digitally as a vector of size N containing N bins. The method used here consists in creating a new vector of the same size N, called $Filter_{median}$. For each component of this vector, the median of the flux is calculated over a time interval, here set at 3 days, which corresponds to a window of a certain number of bins whose size remains fixed throughout the treatment. Then we shift the window by one bin and repeat the operation. The new considered flux, $Flux_{new}$, is calculated from the formula :

\begin{equation}
Flux_{new} = \left( \frac{Flux}{Filter_{median}} - 1 \right)
\end{equation}

Secondly, a threshold was defined in an attempt to remove sharp features due to instrumental effects or transits. Any part of the flux above the threshold is masked.

\subsection{Parameters used for classification}

The features (input parameters) used for classification are the effective temperature (T$_{\rm eff}$) and the luminosity (L) from Gaia DR2 \citep{Gaia1, Gaia2}, and the FliPer, a method developed by \citet{Bugnet18, Bugnet19}, which takes into account the total power in a given band of the PSD. FliPer values are computed using the PSD and an estimation of the photon noise level for each star. Here is the definition of the FliPer, $F_p$, as given by \citet{Bugnet18}:

\begin{equation}
F_p = \overline{PSD} - P_n
\end{equation}

where $\overline{PSD}$ represents the average value of the PSD on a given frequency band and $P_n$ is the photon noise. In this study we consider four FliPer parameters corresponding to four frequency ranges that extend from 0.7, 7, 20 and 50 $\mu$Hz respectively to the Nyquist frequency. Those four FliPer values help to clearly distinguish the shapes of the PSD. 

\begin{figure}[ht!]
 \centering      
 \includegraphics[width=0.48\textwidth,clip]{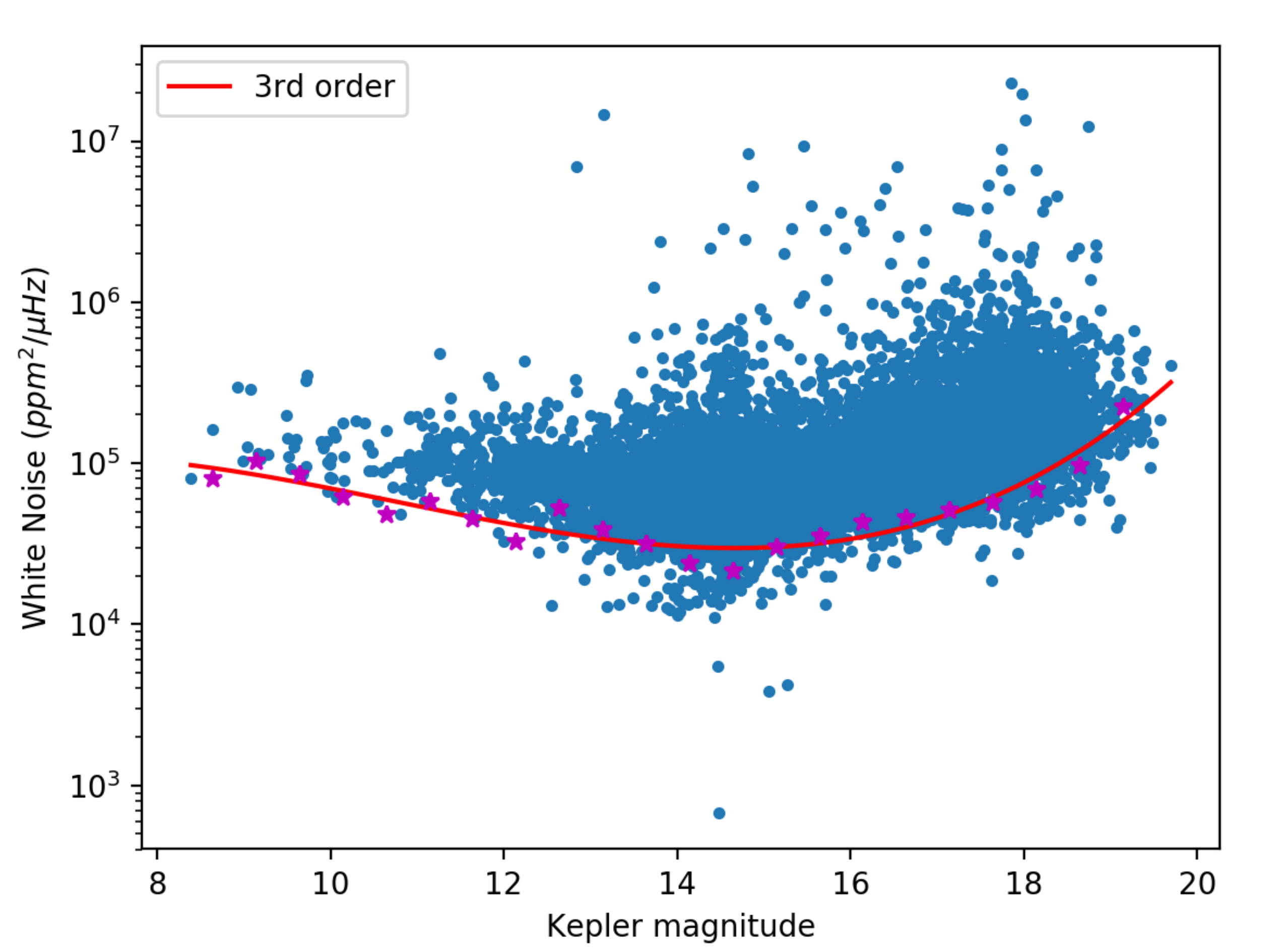}      
  \caption{White noise level for 11 010 stars as a function of their \textit{Kepler} magnitude for Campaign 6. The purple star symbols are an estimation of the bottom of the dense clouds of points and the red line is a 3rd order fit of those purple star symbols.}
  \label{noise}
\end{figure}

Here the method used to determine the photon noise is different from \citet{Bugnet18}. Indeed, the fact that each campaign points a different part of the sky introduces a different noise level in each of those campaigns. So first, we calibrated the photon noise level of each campaign according to the observed \textit{Kepler} magnitude of the star with a method inspired by \citet{Pande18}. For all the stars we computed the mean of the PSD in the range 150 to 280 $\mu$Hz in order to avoid the oscillations frequencies. Then we fitted a 3rd order polynomial function of the noise level at the bottom of the dense cloud of points in bins of width 0.5 magnitude. Points (purple star symbols on Fig. \ref{noise}) that fit the bottom of the cloud of points were adjusted by eye. The white noise level was estimated for stars from campaigns 2 to 8 for now.

\subsection{Machine Learning method}

Facing the huge amount of data we want to classify, the use of an automatic algorithm seems to be the apropriate solution. We decided to use a Random Forest classifier \citep{Breiman01} because of its robustness and its easy implementation. This is a method of supervised learning. That means that the algorithm is trained and validated with a labeled data set. The RF is composed of an ensemble of independent decision trees that gives the most likely class for each star. Each tree is constructed using a random subset of the training set which allows the trees to be independant. To do this the ``RandomForestClassifier" function from the Python Sklearn library \citet{Pedregosa11} was used. \\

At this point, the training and validation sets can be formed. The objective is to separate the stars in four classes : Classical Pulsators (PULS), Main-sequence Solar-Like stars (SL), Red Giants (RG) and Others. The Others category contains all the stars that the classifier could not classify as SL, RG or PULS, including the Eclipsing Binaries. The data set used to train the RF algorithm was constructed from data published on the NASA's website \textit{K2 approved targets \& programs} \footnote{\url{https://keplerscience.arc.nasa.gov/k2-approved-programs.html}}. For the training and validation we dispose of 2,186 labeled stars, for wich the class is already known. This includes 340 SL stars from the originial \textit{Kepler} mission and published by \citet{Chaplin14} because there are only a few SL known among K2 stars. It is possible to add those SL stars from \textit{Kepler} mission because the measure instruments where the same for both missions. But using stars from \textit{Kepler} is not optimal because of the difference in sensitivity that can appear due to the K2 special running mode.  Then we randomly devide this data set in the training set (1,749 stars) and the validation set (379). The \textit{Kepler} stars that were not included in the validation set as they are not stars that we want to classify.

\section{Classification results \& Perspectives}

\begin{figure}[ht!]
 \centering
 \includegraphics[width=0.48\textwidth,clip]{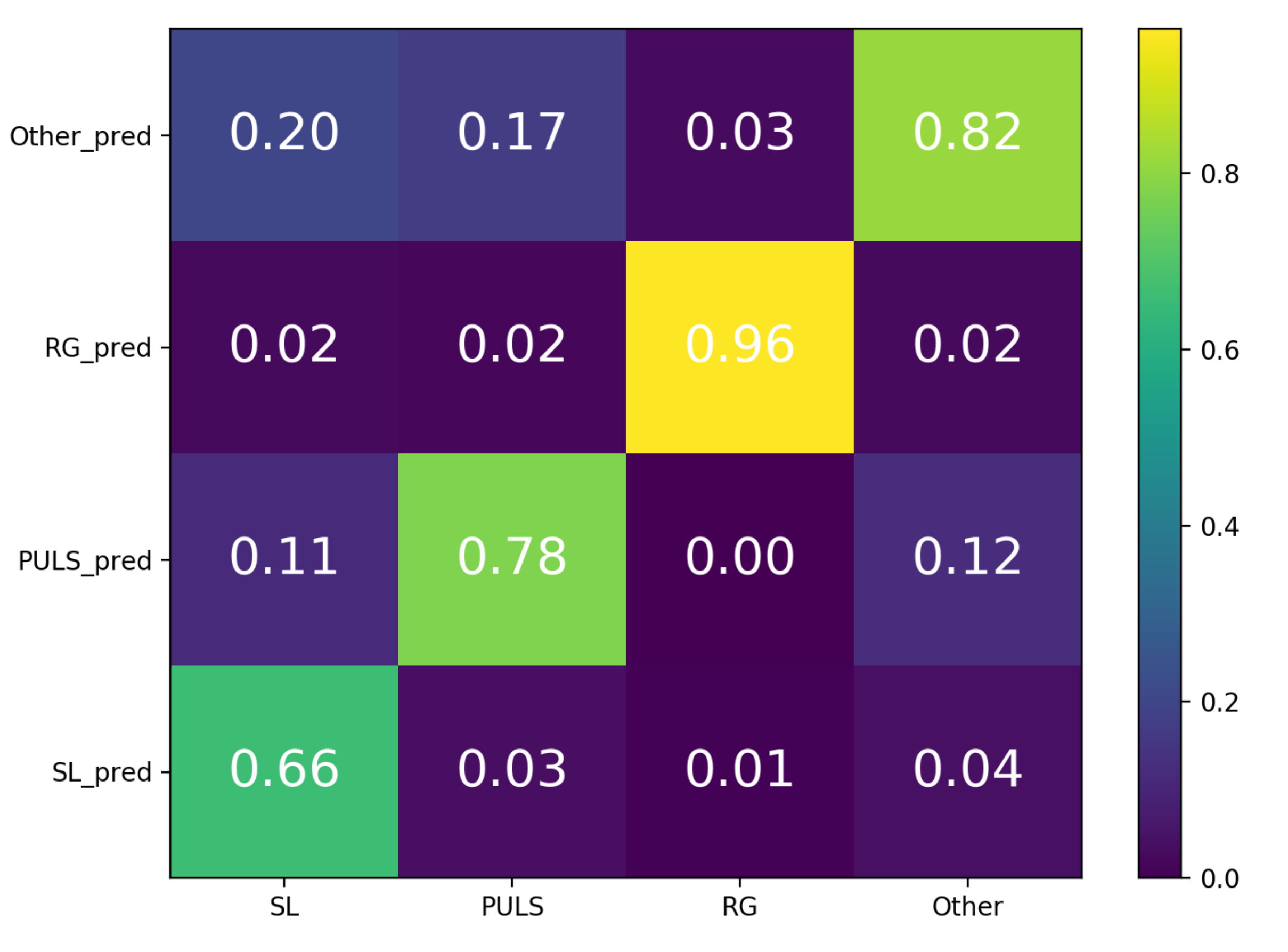}   
  \caption{Confusion matrix resulting from the class estimation of 379 stars from the validation test. The accuracy of the model is 83.0\%. On the diagonal there is the percentage of stars that are well classified. }
  \label{confusion}
\end{figure}

In this work we present a Random Forest method to perform an automatic classification of K2 pulsating stars. Those preliminary results are very encouraging with an accuracy of 83\% on the classification of the validation set wich contain 379 stars. Fig. \ref{confusion} represents the confusion matrix for a classification test. The confusion matrix compares the prediction made by the model with the actual class of a star.  This points out that the main confusion comes from PULS stars that are classified in the Other category (21\%). There are also some SL stars that are predicted as classical pulsators (16\%). On the other hand, we notice that there is not so much confusion between other classes, less than 9\%, and that the RG are very well classified with an accuracy of 98\%. In order to improve those results we plan to increase the number of stars in the training set and improve the lightcurves processing. 

Perspectives for this work are first of all, classify all K2's campaings and then subdivide the PULS category in 7 sub-classes (RR Lyrae, Slowly Pulsating B-type stars, $\delta$-Scuti, $\gamma$- Doradus, $\beta$-Cepheid, Cepheid and rapidly oscillating Ap stars). Finally, this study on K2 data provide a better understanding of how the method works for low-resolution PSDs as it does for most of the millions of lightcurves with a 30 minute cadence from the Full Frame Images that will beprovided by TESS \citep{Ricker14}.

\begin{acknowledgements}
This work was partially fund by PLATO CNES grant. S.M. acknowledges the support from the Ramon y Cajal fellowship number RYC-2015-17697. This paper includes data collected by the K2 mission and obtained from the MAST data archive at the Space Telescope Science Institute (STScI). Funding for the K2 mission is provided by the NASA Science Mission Directorate. STScI is operated by the Association of Universities for Research in Astronomy, Inc., under NASA contract NAS 5-26555.
\end{acknowledgements}

\bibliographystyle{aa}  
\bibliography{LE_SAUX_S15} 

\begin{thebibliography}{16}
\expandafter\ifx\csname natexlab\endcsname\relax\def\natexlab#1{#1}\fi

\bibitem[{{Borucki} {et~al.}(2010){Borucki}, {Koch}, {Basri}, {Batalha},
  {Brown}, {Caldwell}, {Caldwell}, {Christensen-Dalsgaard}, {Cochran},
  {DeVore}, {Dunham}, {Dupree}, {Gautier}, {Geary}, {Gilliland}, {Gould},
  {Howell}, {Jenkins}, {Kondo}, {Latham}, {Marcy}, {Meibom}, {Kjeldsen},
  {Lissauer}, {Monet}, {Morrison}, {Sasselov}, {Tarter}, {Boss}, {Brownlee},
  {Owen}, {Buzasi}, {Charbonneau}, {Doyle}, {Fortney}, {Ford}, {Holman},
  {Seager}, {Steffen}, {Welsh}, {Rowe}, {Anderson}, {Buchhave}, {Ciardi},
  {Walkowicz}, {Sherry}, {Horch}, {Isaacson}, {Everett}, {Fischer}, {Torres},
  {Johnson}, {Endl}, {MacQueen}, {Bryson}, {Dotson}, {Haas}, {Kolodziejczak},
  {Van Cleve}, {Chandrasekaran}, {Twicken}, {Quintana}, {Clarke}, {Allen},
  {Li}, {Wu}, {Tenenbaum}, {Verner}, {Bruhweiler}, {Barnes}, \&
  {Prsa}}]{Borucki10}
{Borucki}, W.~J., {Koch}, D., {Basri}, G., {et~al.} 2010, Science, 327, 977

\bibitem[{{Breiman}(2001)}]{Breiman01}
{Breiman}, L. 2001, Machine Learning, 45, 5

\bibitem[{{Bugnet} {et~al.}(2018){Bugnet}, {Garc\'\i a}, {Davies}, {Mathur},
  {Corsaro}, {Hall}, \& {Rendle}}]{Bugnet18}
{Bugnet}, L., {Garc\'\i a}, R.~A., {Davies}, G.~R., {et~al.} 2018, \aap, 620,
  A38

\bibitem[{{Bugnet} {et~al.}(2019){Bugnet}, {Garc\'\i a}, {Mathur}, {Davies},
  {Hall}, {Lund}, \& {Rendle}}]{Bugnet19}
{Bugnet}, L., {Garc\'\i a}, R.~A., {Mathur}, S., {et~al.} 2019, \aap, 624, A79

\bibitem[{{Chaplin} {et~al.}(2014){Chaplin}, {Elsworth}, {Campantey},
  {Handberg}, {Miglios}, \& {Basus}}]{Chaplin14}
{Chaplin}, W.~J., {Elsworth}, Y., {Campantey}, T.~L., {et~al.} 2014, \mnras,
  445, 946

\bibitem[{{Gaia collaboration et al.}(2016)}]{Gaia1}
{Gaia collaboration et al.} 2016, \aap, 595, A1

\bibitem[{{Gaia collaboration et al.}(2018)}]{Gaia2}
{Gaia collaboration et al.} 2018, \aap, 616, A1

\bibitem[{{Garc\'\i a} {et~al.}(2011){Garc\'\i a}, {Hekker}, {Stello},
  {Guti\'errez-Sotto}, {Handberg}, {Huber}, \& {Karoff}}]{Garcia11}
{Garc\'\i a}, R.~A., {Hekker}, S., {Stello}, D., {et~al.} 2011, \mnras, 414, L6

\bibitem[{{Garc\'{\i}a} {et~al.}(2014){Garc\'{\i}a}, {Ceillier}, {Salabert},
  {Mathur}, {van Saders, J. L.}, {Pinsonneault, M.}, {Ballot, J.}, {Beck, P.
  G.}, {Bloemen, S.}, {Campante, T. L.}, {Davies, G. R.}, {do Nascimento,
  J.-D.}, {Mathis, S.}, {Metcalfe, T. S.}, {Nielsen, M. B.}, {Su\'arez, J. C.},
  {Chaplin, W. J.}, {Jim\'enez, A.}, \& {Karoff, C.}}]{Garcia14}
{Garc\'{\i}a}, R.~A., {Ceillier}, T., {Salabert}, D., {et~al.} 2014, A\&A, 572,
  A34

\bibitem[{{Handberg} \& {Lung}(2014)}]{Handberg14}
{Handberg}, R. \& {Lung}, M.~N. 2014, \mnras, 445, 2698

\bibitem[{Howell {et~al.}(2014)Howell, Sobeck, Haas, Still, Barclay, Mullally,
  Troeltzsch, Aigrain, Bryson, Caldwell, \& Chaplin}]{Howell14}
Howell, S.~B., Sobeck, C., Haas, M., {et~al.} 2014, Publications of the
  Astronomical Society of the Pacific, 126, 398.

\bibitem[{{Luger} {et~al.}(2016){Luger}, {Agol}, {Kruse}, {Barnes}, {Becker},
  {Foreman-Mackey}, \& {Derming}}]{Luger16}
{Luger}, R., {Agol}, E., {Kruse}, E., {et~al.} 2016, \apj, 152, 14 pp.

\bibitem[{{Luger} {et~al.}(2018){Luger}, {Kruse}, {Foreman-Mackey}, {Agol}, \&
  {Saunders}}]{Luger18}
{Luger}, R., {Kruse}, E., {Foreman-Mackey}, D., {Agol}, E., \& {Saunders}, N.
  2018, \apj, 156, 21 pp.

\bibitem[{{Pande} {et~al.}(2018){Pande}, {Bedding}, {Huber}, \&
  {Kjeldsen}}]{Pande18}
{Pande}, D., {Bedding}, T., {Huber}, D., \& {Kjeldsen}, H. 2018, \mnras, 480,
  467

\bibitem[{Pedregosa {et~al.}(2011)Pedregosa, Varoquaux, Gramfort, Michel,
  Thirion, Grisel, Blondel, Prettenhofer, Weiss, Dubourg, Vanderplas, Passos,
  Cournapeau, Brucher, Perrot, \& Duchesnay}]{Pedregosa11}
Pedregosa, F., Varoquaux, G., Gramfort, A., {et~al.} 2011, J. Mach. Learn.
  Res., 12, 2825

\bibitem[{Ricker {et~al.}(2014)Ricker, Winn, Vanderspek, Latham, Bakos, Bean,
  Berta-Thompson, Brown, Buchhave, Butler, \& Butler}]{Ricker14}
Ricker, G.~R., Winn, J.~N., Vanderspek, R., {et~al.} 2014, Transiting Exoplanet
  Survey Satellite (TESS)

\end{thebibliography}

\end{document}